\documentclass[pra]{revtex4}%

\usepackage{graphicx}
\usepackage{dcolumn}
\usepackage{bm}

\begin{document}

\title{Dynamics of tripartite entanglement}

\author{S. Shelly Sharma}
\email[shelly@uel.br]{}
\thanks{}
\affiliation{Depto. de F\'{i}sica, Universidade Estadual de Londrina,
Londrina 86051-990, PR Brazil }

\author{N. K. Sharma}
\email[nsharma@uel.br]{}
\thanks{}
\affiliation{Depto. de Matem\'{a}tica, Universidade Estadual de Londrina,
Londrina 86051-990 PR, Brazil }

\begin{abstract}
Maximally entangled states are of utmost importance to quantum
communication, dense coding, and quantum teleportation. With a trapped ion
placed inside a high finesse optical cavity, interacting with field of an
external laser and quantized cavity field, a scheme to generate a maximally
entangled three qubit GHZ state, is proposed. The dynamics of tripartite
entanglement is investigated, using negativity as an entanglement measure
and linear entropy as a measure of mixedness of a state. It is found that
(a) the number of modes available to the subsystem determines the maximum
entanglement of a subsystem, b) at entanglement maxima and minima, linear
entropy and negativity uniquely determine the nature of state, but the two
measures do not induce the same ordering of states, and c) for a special choice
of system parameters maximally entangled tripartite two mode GHZ state is
generated. The scheme presented for GHZ state generation is a single step
process and is reduction free.

PACS: 03.67.-a, 42.50.-p, 03.67.Dd
\end{abstract}

\maketitle

Maximally entangled states are of utmost importance to quantum communication 
\cite{eker91}, dense coding \cite{benn92}, and quantum teleportation \cite
{benn93}. It is now possible to prepare a cold trapped ion in a given
initial state \cite{wine98} with the trap placed inside a high finesse
optical cavity \cite{schm03}. We propose an experimental scheme, using ion
trap in an optical cavity, to generate a maximally entangled three qubit GHZ
state. For different sets of initial states of the system, entanglement
dynamics of ionic internal state, ionic center of mass state, and cavity
state is investigated using negativity as an entanglement measure and linear
entropy as a measure of state purity.

\section{Tripartite system and interaction Hamiltonian}

Consider a trapped two-level ion in an optical cavity interacting with an
external laser and the quantized cavity field. The quantum state of
tripartite system (ABC) contains information about the internal state of the
ion (sub-system A), the vibrational state of ionic center of mass in trap
(subsystem B) and the state of optical cavity (C). Basis vectors spanning
the Hilbert space of composite tripartite system are $|i,m,n>,$ where $%
|i>,i\in g,e$ represents ion in ground state $(i=g)$ and excited state $(i=e)
$. The vibrational (photonic) number states are denoted by $%
|m>(|n>),m(n)=0,1,2,...\infty $. The Hamiltonian due to interaction of
trapped two-level ion of internal frequency $\omega _{0},$ with resonant
external laser field of frequency $\omega _{L}=\omega _{0},$ and with the
cavity field tuned to red sideband of ionic vibrational motion that is $%
\omega _{0}-\omega _{c}=\nu $, in interaction picture and rotating wave
approximation is given by \cite{shel03}, 
\begin{equation}
\hat{H_{I}}=\hbar \Omega \lbrack \sigma _{+}{\hat{O}_{0}^{L}}+\sigma _{-}{%
\hat{O}_{0}^{L}}]+\hbar g\left[ {\eta _{c}}\sigma _{+}\hat{b}{\hat{O}_{1}^{c}%
}\hat{a}+h.c.\right] .  \label{2.1}
\end{equation}
Here $\hat{a}^{\dagger }(\hat{a})$ and $\hat{b}^{\dagger }(\hat{b})$ are
creation(destruction) operators for vibrational phonon and cavity field
photon, respectively, and $\nu $ is trap frequency. The ion phonon and
ion-cavity coupling constants are $\Omega $ and $g$, whereas $\sigma
_{k}(k=z,+,-)$ are the Pauli operators qualifying the internal state of the
ion. The operator ${{\hat{O}}_{k}\ }$is defined as 
\begin{equation}
{{\hat{O}}_{k}}=\exp \left( -\frac{\eta ^{2}}{2}\right) \sum_{p=0}^{\infty }%
\frac{(i\eta )^{2p}\hat{a}^{\dagger p}\hat{a}^{p}}{p!\left( p+k\right) !},
\label{2.2}
\end{equation}
with Lamb-Dicke (LD) parameters relative to the laser field and the cavity
field denoted by $\eta =\eta _{L}$ and $\eta =\eta _{c}$ respectively. In
the limiting case $\eta _{L}\ll 1$ and $\eta _{c}\ll 1$, ${\hat{O}%
_{k=0,1}\rightarrow 1,}$ as such the relevant part of interaction picture
Hamiltonian reduces to 
\begin{equation}
\hat{H_{I}}=\hbar \Omega \lbrack \sigma _{+}+\sigma _{-}]+\hbar g{\eta _{c}}%
\left[ \sigma _{+}\hat{b}\hat{a}+\sigma _{-}\hat{b}^{\dagger }\hat{a}\right]
.  \label{2.3}
\end{equation}
To obtain unitary time evolution of the system we work in the basis, $\left|
g,m,n\right\rangle $, $\left| e,m,n\right\rangle $, $\left|
g,m-1,n-1\right\rangle ,$ and $\left| e,m-1,n-1\right\rangle $. Eigen values
and eigen vectors of $\hat{H_{I}}$ \ are obtained analytically. Starting
from a given initial state of the system, written in terms of eigen
functions of $\hat{H_{I}},$ the pure state $\Psi (t)$ for the composite
system is obtained by solving the time dependent Schr\"{o}dinger equation, 
\begin{equation}
H_{I}\Psi (t)=i\hbar \frac{d}{dt}\Psi (t).  \label{2.5}
\end{equation}
The density operator for the composite system defined as $\widehat{\rho }%
^{ABC}(t)=\left| \Psi (t)\right\rangle \left\langle \Psi (t)\right| $, can
be used to obtain reduced density operators $\widehat{\rho }^{A}(t),\widehat{%
\rho }^{B}(t),$ and $\widehat{\rho }^{C}(t)$ for subsystems A, B, and C,
respectively.

\begin{figure}[b]
\centering
\includegraphics[width=3.75in,height=5.0in,angle=-90]{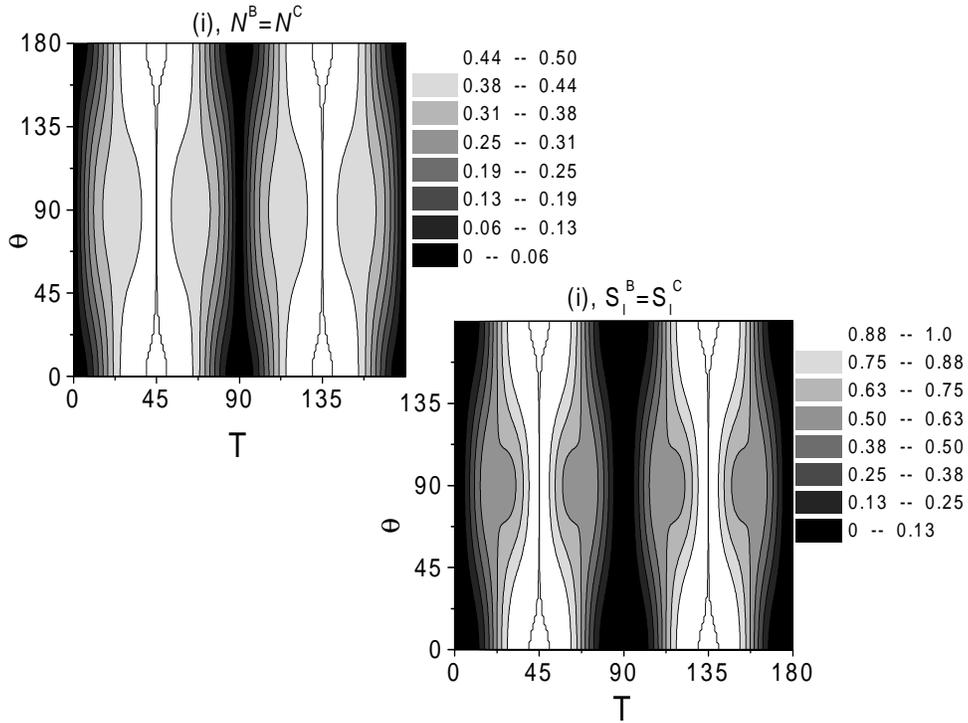}
\caption{Contour plots of negativity $\mathit{N}^{B}(=\mathit{N}^{C}$
 and linear entropy $S_{l}^{B}(=S_{l}^{C})$ as a function of 
$\theta $ and scaled time $T(=at),$ for subsystems $B$ and $C$. 
Initial states are of type (i) that is separable at $T=0$.
 Variables $\theta $ and $T$ are in degrees.}
\label{fig1}
\end{figure}

\section{Initial states of the system and GHZ state generation}

Consider that an ion is prepared initially in state 
\begin{equation}
\Psi (0)=\left( cos(\theta )\left| g\right\rangle +sin(\theta )\left|
e\right\rangle \right) \left| m=0,n=0\right\rangle ,
\end{equation}
with the center of mass in lowest energy trap level ($m=0),$ cavity in it's
vacuum state ($n=0$), and $0\leq \theta \leq \pi $. Defining, $a=\frac{1}{2}%
g\eta _{c}$, and $\mu =\sqrt{a^{2}+\Omega ^{2}},$ we solve Eq.(\ref{2.5})
and verify that for interaction time $t_{p}$ such that $\mu t_{p}=p\pi $, $%
p=1,2,...,$%
\begin{eqnarray}
\Psi (t_{p}) &=&(-1)^{p}\left[ \cos (\theta )\cos (at_{p})\left|
g,0,0\right\rangle +\sin (\theta )\cos (at_{p})\left| e,0,0\right\rangle
\right.   \nonumber \\
&&\left. -i\sin (\theta )\sin (at_{p})\left| g,1,1\right\rangle -i\cos
(\theta )\sin (at_{p})\left| e,1,1\right\rangle \right] .  \label{2.6}
\end{eqnarray}
For the choice $%
{\mu}%
=4a,$ $\theta =q\pi (q=0,1)$ at instant $t_{1}=\left[ \pi /(4a)\right] $ the
system is found to be in state 
\begin{equation}
\Psi (t_{1})=\frac{\left( -1\right) ^{1+q}}{\sqrt{2}}\left( \left|
g,0,0\right\rangle -i\left| e,1,1\right\rangle \right) ,
\end{equation}
which is a maximally entangled GHZ state of the tripartite two mode system.

To get further insight into tripartite entanglement, the time evolution of
entanglement for the following sets of initial states is investigated:

(i) $\Psi (0)=\left( cos(\theta )\left| g\right\rangle +sin(\theta )\left|
e\right\rangle \right) \left| 0,0\right\rangle $ , $0\leq \theta \leq \pi $,
a separable state.

(ii) $\Psi (0)=\left( cos(\theta )\left| g,1\right\rangle +sin(\theta
)\left| e,0\right\rangle \right) \left| 0\right\rangle $ , $0\leq \theta
\leq \pi $, cavity state is separable at $t=0$.

(iii)$\Psi (0)=\left( cos(\theta )\left| g,\beta \right\rangle \ +sin(\theta
)\left| e,-\beta \right\rangle \right) \ \left| 0\right\rangle $ , $0\leq
\theta \leq \pi $, cavity state is separable at $t=0$, with the ion prepared
in a Schrodinger cat state. The coherent state $\left| \beta \right\rangle ,$
is given by

\begin{equation}
\left| \beta \right\rangle =\exp \left( -\frac{\left| \beta \right| ^{2}}{2}%
\right) \sum\limits_{m=0}^{\infty }\frac{\beta ^{m}}{\sqrt{m!}}\left|
m\right\rangle ,
\end{equation}
where $\left| \beta \right| $ is the average number of vibrational quanta
associated with center of mass motion.

\begin{figure}[b]
\centering
\includegraphics[width=3.75in,height=5.0in,angle=-90]{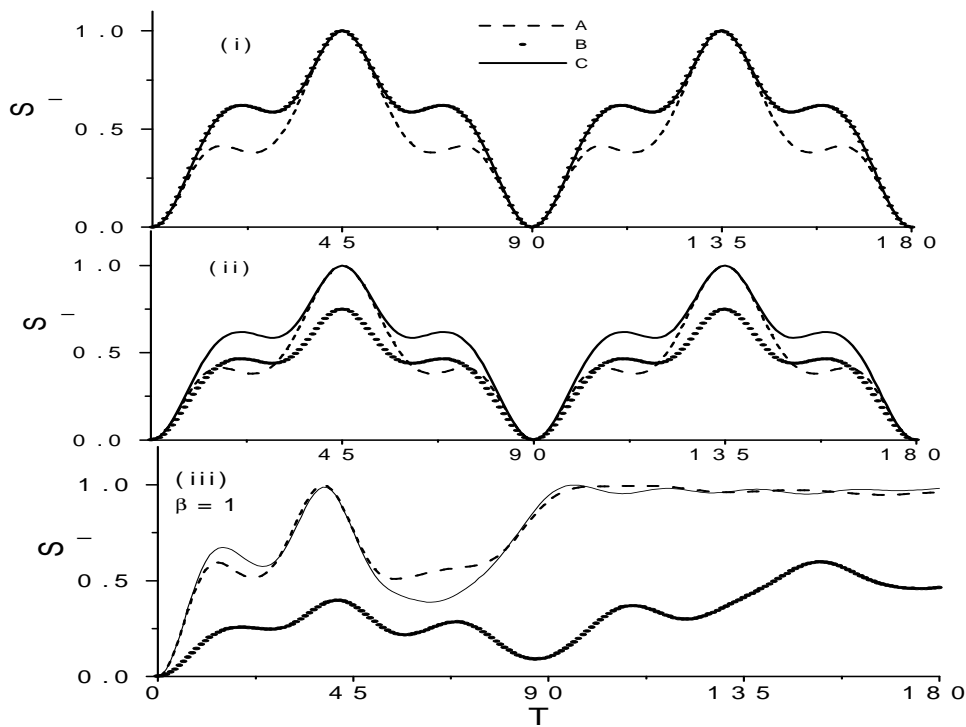}
\caption{ $S_{l}$, versus scaled time $T(=at)$, for initial states of 
the type (i) separable at $T=0$, (ii) cavity state separable at $T=0$, 
and (iii) cavity state separable and ion in a Schrodinger cat state at 
$T=0$, for the choice $\theta =90^{o}$, $\beta=1$.}
\label{fig2}
\end{figure}

\section{ Measures of entanglement and purity of a state}

For a given density matrix operator $\widehat{\rho }^{ABC}(t)$ at instant $%
t, $ acting on composite Hilbert space associated with system ($ABC$),
reduced density operator for subsystem $A$ is defined as, $\widehat{\rho }%
^{A}(t)=Tr_{BC}(\widehat{\rho }^{ABC}(t))$. The partial transpose of density
operator $\widehat{\rho }^{A}$ with respect to subsystem $A$ reads as 
\begin{equation}
\widehat{{\rho }}^{T_{A}}=\sum_{i,j=1}^{d_{A}}\sum_{m,r}^{d_{B}}%
\sum_{n,s}^{d_{C}}\left\langle i,m,n\left| \widehat{{\rho }}^{A}\right| \
j,r,s\right\rangle \left| \ j,m,n\right\rangle \left\langle \ i,r,s\right| .
\label{3.3}
\end{equation}

Negativity \cite{vida02} for subsystem $A$ defined in terms of trace norm
of\ partial transpose of density matrix as 
\begin{equation}
\mathit{N}^{A}=\frac{\left\| \rho ^{T_{A}}\right\| -1}{2},
\end{equation}
is equal to the modulus of sum of negative eigenvalues of operator $\widehat{%
{\rho }}^{T_{A}}.$ $\mathit{N}^{A}$ is a measure of entanglement of
subsystem $A$ with it's complement ($BC$) in the composite quantum system ($%
ABC$). Measures of entanglement, $\mathit{N}^{B}$ and $\mathit{N}^{C},$ for
subsystems $B$ and $C$ with their respective complementary systems ($AC$)
and ($AB$) are defined in analogous manner.

Although matrix $\rho ^{ABC}(t)$ represents a pure state, the reduced
density matrices, $\rho ^{A}(t),\rho ^{B}(t)$ and $\rho ^{C}(t)$ do not
necessarily do so. Linear Entropy, $S_{l}$ defined as 
\begin{equation}
S_{l}=\frac{d}{d-1}\left( 1-tr(\rho ^{2})\right) ,
\end{equation}
is used as a measure of purity of a state. For a pure state $S_{l}=0$ while
for a maximally mixed state $S_{l}=1.$ Here $d$ is the number of modes
available to a subsystem.

\begin{figure}[t]
\centering
\includegraphics[width=3.75in,height=5.0in,angle=-90]{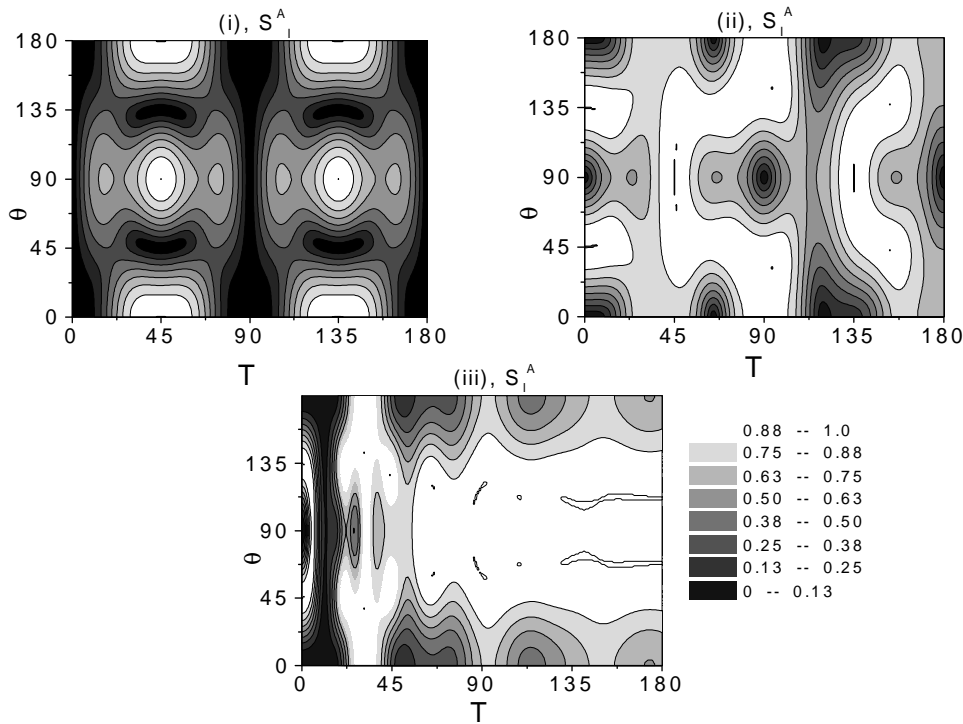}
\caption{Contour plots of linear entropy $S_{l}^{A},$ as a function of $%
\theta $ and $T(=at)$ , for initial states of the types (i) separable at 
$T=0$, (ii) cavity state separable at $T=0,$ and (iii) cavity state separable 
and ion in a Schrodinger cat state ($\beta=1$) at $T=0$.}
\label{fig3}
\end{figure}

\section{Results and conclusions}

For the three sets of states described in section (2), we have obtained
analytic expressions for $\Psi ^{ABC}(t)$. These expressions have been used
for numerical calculations of entanglement measures $\mathit{N}^{A},\mathit{N%
}^{B}$ and $\mathit{N}^{C},$ for the system parameter choice of $%
{\mu}%
/a=4$. We have also calculated the reduced density matrices, $\rho
^{A}(t),\rho ^{B}(t),$ $\rho ^{C}(t)$ and from these the linear entropies $%
S_{l}^{A},S_{l}^{B}$ and $S_{l}^{C}$. Some of the results are displayed in
Figs. (1) to (3). Fig. (1) shows contour plots of negativity $\mathit{N}%
^{B}(=\mathit{N}^{C})$ and linear entropy $S_{l}^{B}(=S_{l}^{C})$ as a
function of $\theta (0\leq \theta \leq \pi ).$ and scaled time $T(=at).$
Initial states are separable at $T=0$ (type (i)). A comparison of contour
shapes denounces the fact that negativity and linear entropy do not produce
the same ordering \ of states. However at zero and maximum entanglement, the
states correspond to pure states and maximally mixed states, respectively.
We find the subsystems A and B to be maximally entangled and maximally mixed
at $T=45^{o}$ and $135^{o}$ for all possible choices of $\theta .$

Fig (2) displays for the choice $\theta =90%
{{}^\circ}%
,$ linear entropy $S_{l}$ versus scaled time $T(=at)$ in degrees, for
subsystems $A$, $B$ and $C$. Parts (i), (ii) and (iii) in the figure refer
to initial states of the type (i), (ii), and (iii), as in section (2). For
initial state (i), all three subsystems are found to be in maximally mixed
state at $T=45^{o}$and $135^{o}$, hence are maximally entangled. For initial
state (ii), although subsystems $A$ and $C$ become maximally entangled at $%
T=45^{o}$and $135^{o}$, system $B$ fails to reach maximum entanglement. For
case (iii), when ion's vibrational state is a coherent state ($\beta =1$) at 
$t=0$, entanglement is seen to increase with time for all the three
subsystems. As expected, the inial state temperature has strong influence on
generation of maximally entangled tripartite state.

Contour plots of linear entropy $S_{l}^{A},$ as a function of $\theta $ and $%
T(=at)$ , for initial states of the types (i), (ii), and (iii) are displayed
in Fig. (3). For initial states (i) subsystem $B$ is a qubit, for initial
states (ii) subsystem $B$ is a qutrit, and for initial states (iii) a large
number of modes are available to system $B$, while $A$ and $C$ remain as two
mode systems. A comparison of parts (i), (ii), and (iii) of Fig. (3) shows
that for a given choice of $\theta $, the time evolution of purity and
entanglement of subsystem $A$ depends strongly on the number of modes
available to subsystem $B$. Time evolution of $S_{l}$ for system $A$ in
Figure (2) is a particular case of Fig. (3). We conclude that (a) the number
of modes available to a subsystem not only determines the maximum entanglement of that 
subsystem but also has a strong influence on entanglement of other subsystems, 
b) at entanglement maxima and minima, linear entropy and
negativity uniquely determine the nature of state, but the two measures do
not induce the same ordering of states, c) for a special choice of system
parameters maximally entangled tripartite two mode GHZ state is generated.
The scheme presented for GHZ state generation is a single step process and
is reduction free.

{\Large Acknowledgments}

Financial Support from Funda\c{c}\~{a}o Arauc\'{a}ria, Pr, Brazil is
acknowledged.

\newpage{}

\end{document}